\documentclass[prl,showpacs,twocolumn]{revtex4}
\usepackage{amssymb}
\usepackage{amsmath}
\usepackage{graphicx}

\usepackage{color}
\definecolor{blue}{rgb}{0,0,1}
\definecolor{grey}{rgb}{0.6,0.6,0.6}

\def    \bse{\begin{subequations}}
\def    \ese{\end{subequations}}

\def \be{\begin{equation}}
\def \ee{\end{equation}}
\def \bew{\begin{widetext}\begin{equation}}
\def \eew{\end{equation}\end{widetext}}
\def \bmlett{\begin{mathletters}}
\def \emlett{\end{mathletters}}

\def \ha{\hat{a}}
\def \hb{\hat{b}}
\def \hd{\hat{d}}
\def \hH{\hat{H}}
\def \hd{\hat{d}}

\def \omegam{\omega_M}

\def \omegam{\omega_{\rm M}}
\def \Gammam{\gamma}

\def \bhB {\hat{\beta}_{B}}
\def \bhA {\hat{\beta}_{A}}
\def \nthA {\bar{n}_{\mathrm{th},A}}

\def \nthone {\bar{n}_{\mathrm{th},1}}
\def \nthtwo {\bar{n}_{\mathrm{th},2}}

\def \Gco {\Gamma_{\rm opt}} 

\begin{document}

\title{Reservoir-engineered entanglement in optomechanical systems}
\author{Ying-Dan Wang}
\affiliation{Department of Physics, McGill University, 3600 rue
University, Montreal, QC Canada H3A 2T8}
\author{Aashish~A.~Clerk}
\affiliation{Department of Physics, McGill University, 3600 rue
University, Montreal, QC Canada H3A 2T8}

\date{June 18, 2013}

\begin{abstract}
We show how strong steady-state entanglement can be achieved in a three-mode 
optomechanical system (or other parametrically-coupled bosonic system) by effectively laser-cooling a delocalized Bogoliubov mode. This approach allows one to surpass the bound on the maximum stationary intracavity entanglement possible with a coherent two-mode squeezing interaction. In particular, we find that optimizing the relative ratio of 
optomechanical couplings, rather than simply increasing their magnitudes, is essential for achieving strong entanglement.  
Unlike typical dissipative entanglement schemes, our results cannot be described by treating the effects of the entangling
reservoir via a Linblad master equation.
\end{abstract}


\maketitle



\textit{Introduction-- }
The study of highly entangled quantum states is of interest both for fundamental reasons and for a myriad of applications to quantum information
processing and quantum communication. Of particular fundamental interest is the possibility to entangle distinct macroscopic objects, a task made difficult by the 
unavoidable decoherence and dissipation associated with such systems. Equally interesting would be the ability to entangle photons of very different frequencies, e.g. microwave and optical photons.    

A promising venue for the realization of both these kinds of entanglement is provided by quantum optomechanics, where macroscopic mechanical degrees of freedom can be controlled, measured and coupled using the modes of an electromagnetic cavity.  Recent milestones in this field include the ability to cavity-cool a mechanical resonator to its ground state of motion \cite{Teufel2011b,Painter2011b} and the observation of many-photon strong coupling effects~\cite{Kippenberg2010b,SafaviNaeini2011b,Teufel2011,Kippenberg2012}.  
A natural setting for entanglement generation is a three-mode optomechanical system consisting of two ``target" modes to be entangled, which are each coupled to a third  ``auxiliary" mode. 
One could either have two optical target modes and a mechanical auxiliary mode, or vice-versa; both variants have recently been achieved in experiment \cite{Painter2012,Wang2012,Sillanpaa2012}.
Several theoretical studies have described such schemes, using the basic idea that the auxiliary mode mediates an effective (coherent) two-mode squeezing interaction between the two target modes, see e.g. \cite{Mancini2002,Vitali2007c,Vitali2011,Vitali2012}. However, such schemes typically yield at best a relatively small amount of intra-mode entanglement (something which we quantify more fully below).  

In this paper, we again consider generating steady-state entanglement of two bosonic modes in a three-mode system; while we focus on an optomechanical realization,
our ideas could also be realized using superconducting circuits coupled via Josephson junctions \cite{Devoret2010b,Deppe2012} or other parametrically-coupled 3-bosonic-mode system.
Unlike previous works, we consider the possibility of entanglement via reservoir engineering~\cite{Zoller1996}:  we wish to tailor the dissipative environment of the two target modes such that the dissipative dynamics relaxes the system into an entangled state. Such dissipative entanglement has been discussed in the context of atomic systems~\cite{Plenio2002,Cirac2004,Parkins2006,Zoller2008,Polzik2011b} and has even been realized experimentally~\cite{Polzik2011}. 

The dissipative entanglement scheme we describe is related to optomechanical cavity-cooling schemes \cite{Marquardt07,WilsonRae07} which have been used successfully to cool mechanical resonators to the ground state. In our case, one is not cooling a simple mechanical mode to the ground state, but rather a
hybrid mode delocalized over both target modes. In contrast to previous reservoir-engineering approaches to entanglement generation, where the dynamics
is reduced to a simple Markovian master equation for the target degrees-of-freedom, our treatment is valid even in the regime where a simple adiabatic elimination of
the intermediate mode is not possible.
As we show, this regime turns out to be the most effective at generating entanglement. Our result shows that by optimizing the ratio of optomechanical coupling strengths, rather than simply increasing their magnitudes, this laser-cooling mechanism can be used to yield large amounts of time-independent 
intra-cavity entanglement.  The amount of entanglement is far greater than in previous studies and, in fact, far greater than the maximum possible entanglement allowed by a coherent parametric interaction.
Note that reservoir-engineering in optomechanics has previously been studied theoretically, with the very different goal of generating long-range coherence in arrays~\cite{Tomadin2012}.

\begin{figure}[tp]
\begin{center}
\includegraphics[width= 0.70 \columnwidth]{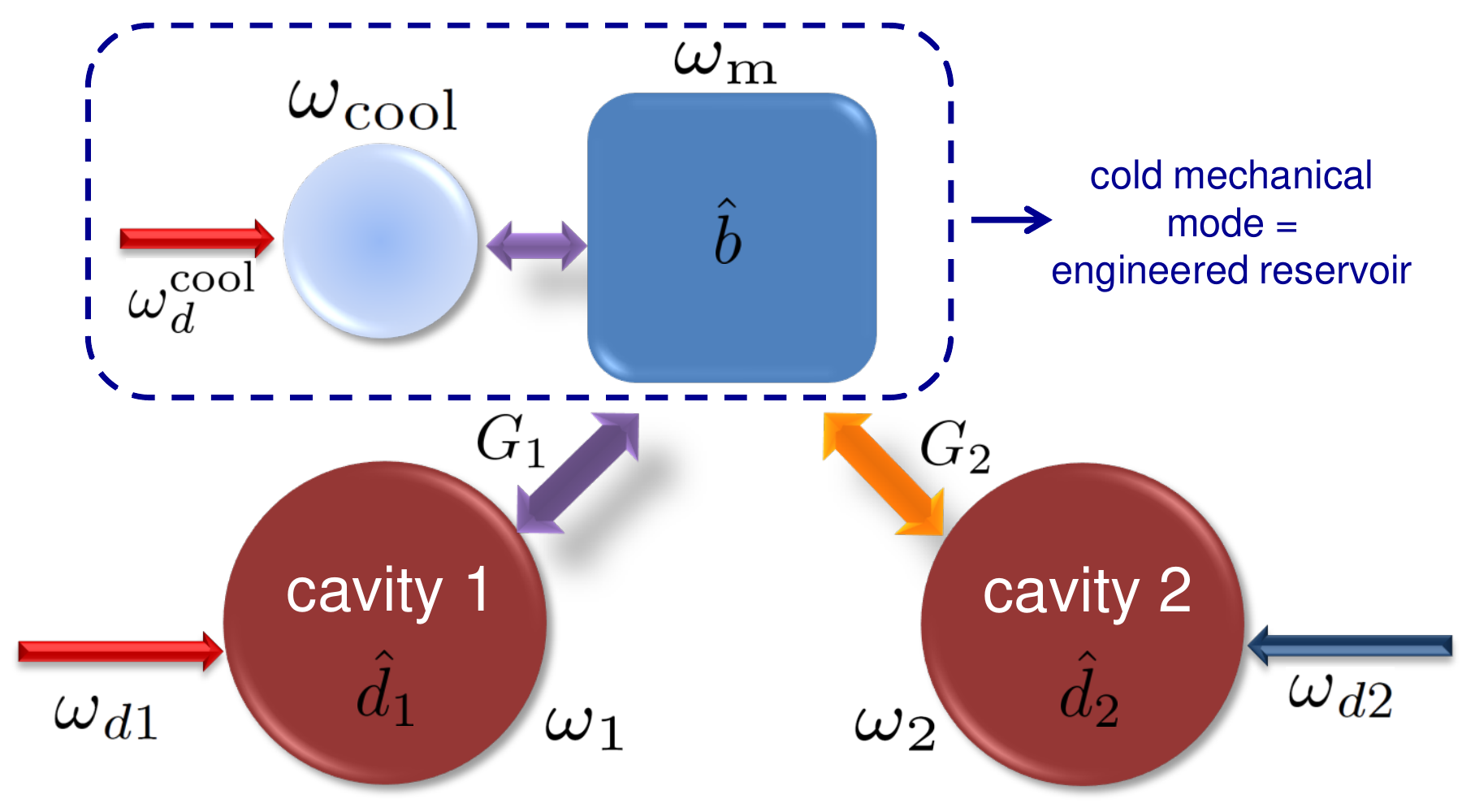}
\end{center}
\vspace{-0.5cm} 
\caption {Schematic of one realization of a 3-mode optomechanical system, where two cavity modes are coupled to a single mode 
of a mechanical resonator.  By driving cavity $1$ ($2$) at the
red (blue) detuned mechanical sideband, a dissipative entanglement mechanism is realized.  To enhance the scheme, the mechanical resonator is cavity cooled and optically damped
via a coupling to a third driven cavity mode, so that its total damping rate $\gamma$ is greater than the cavity damping rates $\kappa$.}
\label{fig:schematics}
\end{figure}


\textit{System and normal modes--}
While our scheme applies to a general bosonic three mode system, we focus here on an optomechanical system where two
optical or microwave cavity modes are coupled to a single mode of a mechanical resonator (see Fig.~1); see the 
EPAPS \cite{EPAPS2013a} for a discussion of entangling two mechanical modes coupled to a cavity mode.
The Hamiltonian is:
\begin{equation}
	\hH=\omegam  \hb^\dag \hb
			+ \sum_{i=1,2}  \left( \omega _{i}  \ha_{i}^{\dag }\ha_{i} + 
			g_{i}\left(  \hb^{\dag }+\hb \right)  \ha_i^\dag \ha_i \right) + \hH_{\rm diss}.
\end{equation}
$\ha_{i}$ is the annihilation operator for cavity $i$ (frequency $\omega _{i}$,  damping rate $\kappa _{i}$), 
$\hb$ is the annihilation operator of the mechanical mode (frequency $\omegam$, damping rate $\Gammam$), and $g_{i}$ are the 
optomechanical coupling strengths.  $\hH_{\rm diss}$ describes the dissipation of each mode, as well as the driving of the cavity modes.
To achieve an entangling interaction, cavity $1$ ($2$) is driven at the red (blue) sideband associated with the mechanical resonator:
$\omega_{d1} = \omega_1 - \omegam$ and $\omega_{d2} = \omega_2 + \omegam$ \cite{Vitali2011}.  We work in an interaction picture with respect to 
the cavity drives,
and write $\ha_i = \bar{a}_i + \hd_i$ where $\bar{a}_i$ is the classical
cavity amplitude.  We take $ |\bar{a}_{1,2}| \gg 1$, which allows us to linearize the optomechanical interaction in the usual way 
(i.e.~drop interaction terms not enhanced by the classical cavity amplitudes).  The linearized Hamiltonian in the rotating frame is thus 
$\hH =  \omega_M \left(\hb^\dag \hb + \hd_1^\dag \hd_1 - \hd_2^\dag \hd_2 \right)+ \hH_{\rm int} + \hH_{\rm CR} + \hH_{\rm diss} $ with
\begin{eqnarray}
		\hH_{\rm int} &=&					
			G_{1}\left(  \hb^{\dag } \hd_{1} + \hd_{1}^{\dag }\hb  \right)
					+G_{2}\left( \hb \hd_{2} +\hd_{2}^{\dag }\hb^{\dag }\right), 
							\label{eq:Hint}		\\
		\hH_{\rm CR} &=&					
			G_{1}\left(  \hb^{\dag } \hd^\dag _{1} + \hd_{1} \hb  \right)
					+G_{2}\left( \hb^\dag \hd_{2} +\hd_{2}^{\dag }\hb \right).
		\label{eq:HCR}
\end{eqnarray}
Here $G_i = g_i \bar{a}_i$ (we take $g_i, \bar{a}_i > 0$ without loss of generality). 
We further focus on the resolved-sideband regime
$\omegam  \gg \kappa_1, \kappa_2$, which suppresses the effects of the non-resonant interactions in $\hH_{CR}$.  The remaining interaction $\hH_{\rm int}$ in Eq.~(\ref{eq:Hint}) has the basic form suitable for entangling $\hd_1$ and $\hd_2$: on a heuristic level, the parametric-amplifier interaction ($G_2$ term) first entangles  $\hd_2$ and $\hb$, and then the beam-splitter interaction ($G_1$ term) swaps the $\hb$ and $\hd_1$ states, thus yielding the desired entanglement.

Note that if one made the interactions in Eq.~(\ref{eq:Hint}) non-resonant (e.g.~by detuning the cavity drives from the sideband resonances by $\Delta$), one
could adiabatically eliminate the mechanical mode, resulting in a two-mode squeezing interaction $\hH_{\rm TMS} \simeq \lambda \left(\hd_1 \hd_2 + {\rm h.c. } \right)$ with 
$\lambda \sim G_1 G_2 / \Delta$~\cite{Marquardt2012}. Such an interaction naturally leads to entanglement, but the amount is severely limited by the requirement of stability $\lambda \leq \kappa_{1,2}/2$. We quantify the entanglement using the standard measure of the logarithmic negativity $E_N$ \cite{EPAPS2013a}.  One finds that the maximum stationary intracavity entanglement due to the two-mode squeezing coupling is (for $\kappa_1 = \kappa_2$ and zero temperature) $E_N = \ln (1 + 2 \lambda / \kappa) \leq \ln 2 \sim 0.7$. 
Many suggested schemes for entanglement generation in optomechanical systems are limited by this stability requirement.

In contrast, the resonant case we consider allows for an alternative dissipative entanglement mechanism capable of much larger $E_N$. We 
will focus attention on the regime $G_2 < G_1$ where (for $\kappa_1 = \kappa_2$) our linear system is always stable \cite{EPAPS2013a}.  
Defining the effective two-mode squeezing parameter $r = \mathrm{ arctanh } (G_2/G_1)$, we introduce delocalized (canonical) cavity Bogoliubov mode operators: 
\begin{eqnarray}
		\bhA &=&
				\hd_{1}\cosh r+\hd_{2}^{\dag }\sinh r\equiv 
					\hat{S}\left( r\right) \hd_{1} \hat{S}^{\dag }\left( r\right),  
						\nonumber \\
		\bhB &=&
				\hd_{1}^{\dag }\sinh r+\hd_{2}\cosh r \equiv \hat{S}\left( r\right) \hd_{2} \hat{S}^{\dag}\left( r\right).   \label{eq:BogModes}
\end{eqnarray}
Here, $\hat{S}\left( r\right) \equiv \exp \left[r \hd_{1} \hd_{2}- h.c. \right] $ is a two-mode squeezing operator.   It thus
follows that the joint vacuum of $\bhA, \bhB$ is the two-mode squeezed state $|r\rangle  = \hat{S}(r) |0,0\rangle$, 
where $|0,0 \rangle$ is the vacuum of $\hd_1, \hd_2$. The entanglement of this state is simply $E_N = 2 r$.  

In terms of these new operators, $\hH_0 = \omegam \left( \hb^\dag\hb + \bhA^\dag \bhA - \bhB^\dag \bhB\right)$ and the optomechanical interactions in Eq.~(\ref{eq:Hint}) and 
(\ref{eq:HCR}) take the simple form
\begin{eqnarray}
	\hH_{\rm int} = \tilde{G} \bhA^\dagger \hb + h.c.,
	\;\;\;\;
	\hH_{\rm CR} = \tilde{G} \bhA^\dagger \hb^\dagger + h.c.,
	\label{eq:HintBog}
\end{eqnarray}
where $\tilde{G} \equiv \sqrt{G_1^2 - G_2^2}$.  The mode $\bhB$ completely decouples from the mechanics (it is a mechanically-dark mode~\cite{Clerk2012a,Tian2012}), while
in the good-cavity limit of interest $\hH_{\rm CR}$ can be neglected, implying that the mode $\bhA$ has a simple beam-splitter interaction
with the mechanics.  $\hH_0 + \hH_{\rm int}$ is trivially diagonalized,  resulting in hybridized modes $\hat{\beta}_{\pm} = \left( \bhA \pm \hb \right) / \sqrt{2}$ with energies 
$\omegam \pm \tilde{G}$.  The
existence of three distinct eigenmodes  (two hybrid, one dark) can be useful to understand entanglement (in particular spectral entanglement~\cite{Vitali2012, Nergis2008}) 
in the case where the mechanical mode is driven by excessive thermal noise; we will 
discuss this in a future work.  
We focus here on generating intracavity entanglement, which has the benefit of being insensitive to whether internal losses contribute to the 
damping rate $\kappa$ of the cavities.

We now exploit the fact that Eq.~(\ref{eq:HintBog}) has exactly the form used for standard cavity cooling \cite{Marquardt07,WilsonRae07}.  
Thus, if we can couple the mechanical
mode $\hb$ to a cold reservoir, then the beam-splitter coupling $\hH_{\rm int}$ can be used to cool $\bhA$ towards vacuum, resulting in a {\it stationary} entangled state.  A high-frequency, low-$Q$
mechanical resonator would thus be ideal.   Alternatively,  we will take the mechanical mode to be coupled to a third cavity mode which is used to laser cool its thermal occupancy towards the ground state by
providing a source of cold damping, see Fig.~1.  In what follows, we include the cooling cavity coupled to the mechanical resonator in the definition of its effective thermal bath; hence, the damping rate $\gamma$ includes the large contribution of the optical damping.  Amusingly, our scheme is one of the few examples in optomechanics where the enhanced mechanical damping rate resulting from cavity cooling is actually highly beneficial.

\begin{figure}[tp]
\begin{center}
\includegraphics[width= 0.9\columnwidth]{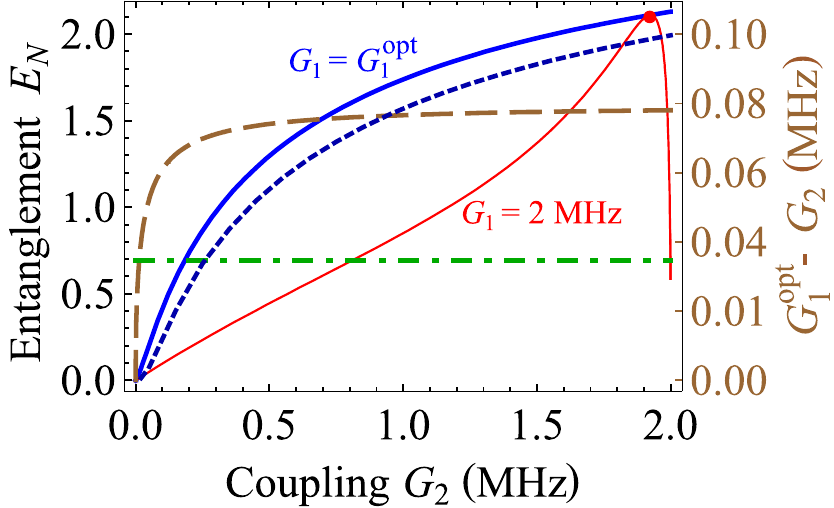}
\end{center}
\vspace{-0.5cm} 
\caption {Stationary intra-cavity entanglement (as quantified by log negativity $E_N$, left scale) as a function of the entangling interaction $G_2$.
We take $\bar{n}_{{\rm th},1} = \bar{n}_{{\rm th},2} = 0$, a mechanical frequency of $\omegam = 2 \pi \times 10 \text{ MHz}$ and damping
$\gamma = 2 \pi \times 0.8 \text{ MHz}$, and {\it do not} make the rotating wave approximation.
The solid red thin curve corresponds to a fixed value of $G_1 = 2 \pi \times 2 \text{ MHz}$, and $\kappa_1 = \kappa_2 = 2 \pi \times 50 \text{ kHz}$. One clearly
sees a non-monotonic dependence on $G_2$. The solid blue thick curves and short-dashed blue curves instead correspond to tuning $G_1$ to the value $G_1^\mathrm{opt}$
for each $G_2$, such that the dissipative entangling mechanism can be optimized. The value of $G_1^\mathrm{opt}-G_2$ (c.f.~Eq.~(\ref{eq:OptimalGTilde})) is indicated by the long-dashed brown curve (right scale). The solid blue (dashed blue) corresponds to $\kappa_1 = \kappa_2 = 50 \text{ kHz}$ ($\kappa_{1/2} / 2 \pi = 45 \text{ kHz}$, $55 \text{ kHz}$) and $\bar{n}_{\rm th, M} = 0$ ($\bar{n}_{\rm th, M} = 0.3$).  The green dash-dot line represents the maximum stationary entanglement achievable with a coherent two-mode squeezing interaction, $E_N = \ln 2$. } 
\label{fig:Figure2}
\end{figure}

\textit{Langevin equations \& cavity cooling--}
To describe the cooling potential of $\hH_{\rm int}$,  we next use input-output theory 
to derive the Heisenberg-Langevin equations for our linearized system.  These take the standard form:
\begin{eqnarray}
		\frac{d}{dt}\hb &=&
				\left(-i \omegam - \frac{\Gammam }{2} \right) \hb  -i \left( G_{1} \hd_{1}
				+G_{2} \hd_{2}^{\dag}\right)- \sqrt{\Gammam} \hb_{\mathrm{in}}, \nonumber \\
		\frac{d}{dt}\hd_{1} &=&
				\left(-i \omegam -\frac{\kappa _{1}}{2} \right) \hd_{1}-iG_{1}\hb - \sqrt{\kappa _{1}}\hd_{1,\mathrm{in}}, \nonumber \\
		\frac{d}{dt}\hd_{2}^{\dag } &=&
				\left(-i \omegam -\frac{\kappa_{2}}{2} \right) \hd_{2}^{\dag }+iG_{2} \hb - \sqrt{\kappa _{2}}\hd_{2,\mathrm{in}}^{\dag},
		\label{eqs:Langevn}
\end{eqnarray}
where $\hd_{i, \rm in}, \hb_{\rm in} $ describes operator-valued white noise driving the cavity and mechanical modes, and we have taken the good-cavity
limit $\omegam \gg \kappa$ (allowing us to drop terms due to $\hH_{\rm CR}$).
Eqs.~(\ref{eqs:Langevn}) are readily solved to find the steady-state occupancy and correlation of the Bogoliubov modes.  In the following analytic expressions, we take $\kappa_1 = \kappa_2$ for simplicity and focus on the good-cavity limit (though Fig.~\ref{fig:Figure2} includes corrections due to $\hH_{\rm CR}$).

Imagine first that the optomechanical interactions vanished, i.e. $\hH_{\rm int} = 0$, and consider the behaviour of 
$\bhA, \bhB$ (defined for a fixed $r > 0$).
Even at zero temperature, 
$\bhA$ and $\bhB$ will have a non-zero
occupancy:  the Bogoliubov transformation of Eq.~(\ref{eq:BogModes}) implies that vacuum noise driving the cavities acts as effective thermal noise for $\bhA,\bhB$.  
Writing these intrinsic ($\hH_{\rm int} = 0$) occupancies as 
$\left	\langle 	\hat{\beta}_j^\dagger 	\hat{\beta}_j  \right\rangle_0  =  \bar{n}_{ {\rm th}, j}$ we have
\begin{eqnarray}
				\bar{n}_{ {\rm th}, A/B} = 
					\bar{n}_{ {\rm th}, 1/2} \cosh ^{2}r +
					\left( \bar{n}_{ {\rm th}, 2/1}+1\right) \sinh ^{2}r. 
 			 \label{n00f} 
\end{eqnarray}
Here, $\nthone$ ($\nthtwo$) represents the temperature of the thermal bath coupled to mechanical resonator $1$ ($2$).  
As one increases the squeeze parameter $r$, the effective heating of the $\hat{\beta}_j$ modes becomes exponentially large, implying the state of the system is far from being
an ideal two-mode squeezed vacuum state; the state is not entangled.

Including now the effects of $H_{\rm int}$ (and taking $r = \mathrm{ arctanh } (G_2/G_1)$), the dark-mode $\bhB$ is unaffected, whereas the occupancy of $\bhA$ is modified to
\begin{eqnarray}
		\left \langle \bhA^{\dag } \bhA\right\rangle & = &
				\frac{\kappa}{  \Gco + \kappa}
						\left(1 + \frac{  \Gco  }{ \Gammam+\kappa  } \right) \nthA + \nonumber \\
						&&
				\frac{\Gco \gamma}{ ( \Gco + \kappa) (\gamma + \kappa)}
					\bar{n}_{\rm th, M}.
		\label{eq:BetaACooling}
\end{eqnarray}
where $\bar{n}_{\rm th, M}$ represents the temperature of the mechanical bath (which includes the cooling cavity), and 
the effective ``cold damping rate" of $\bhA$ by the mechanics is $\Gco\equiv 4\tilde{G}^{2}/\Gammam$.
This is the familiar equation for cavity cooling in the good-cavity limit, where now the mechanics plays the role of a cold reservoir.
For $\bar{n}_{\rm th, M}=0$ and weak coupling ($\Gco \ll \Gammam$), the $\bhA$ mode is cooled by a factor
$\kappa / (\kappa + \Gco)$.
In the strong coupling limit, the cooling factor saturates to a value $\kappa / (\gamma + \kappa)$.

Thus, while even vacuum noise tends to heat $\bhA$, $\bhB$ to an exponentially-large effective temperature, the optomechanical interaction
of Eq.~(\ref{eq:HintBog}) can be used to cool $\bhA$. Using the inequality of Duan et al. \cite{Duan2000}, 
one can show that if one cools $\bhA$ so that
\begin{equation}
	\langle \bhA^\dag \bhA \rangle \leq \sinh^2 r , 
	\label{eq:Duan}
\end{equation}	
then the two cavities {\it must necessarily} be entangled~\cite{EPAPS2013a}.  As the orthogonal Bogoluibov mode $\bhB$ is decoupled from the mechanics, it is not cooled, 
making it impossible to achieve an ideal two-mode squeezed vacuum state. Nonetheless, we find that simply cooling 
$\bhA$ \emph{is} sufficient to generate a steady state with significant entanglement ($E_N \sim 2 r - \ln 2 $ in the large $r$ limit); this is despite the fact that the resulting state has negligible overlap
with a two-mode squeezed vacuum (see EPAPS~\cite{EPAPS2013a}).

To rigorously quantify the cavity-cavity entanglement, we compute and discuss in what follows the log negativity $E_N$, which is a function of the covariance matrix; details are provided in the supplementary material \cite{EPAPS2013a}.


\begin{figure}[tbp]
\begin{center}
\includegraphics[width=0.95\columnwidth]{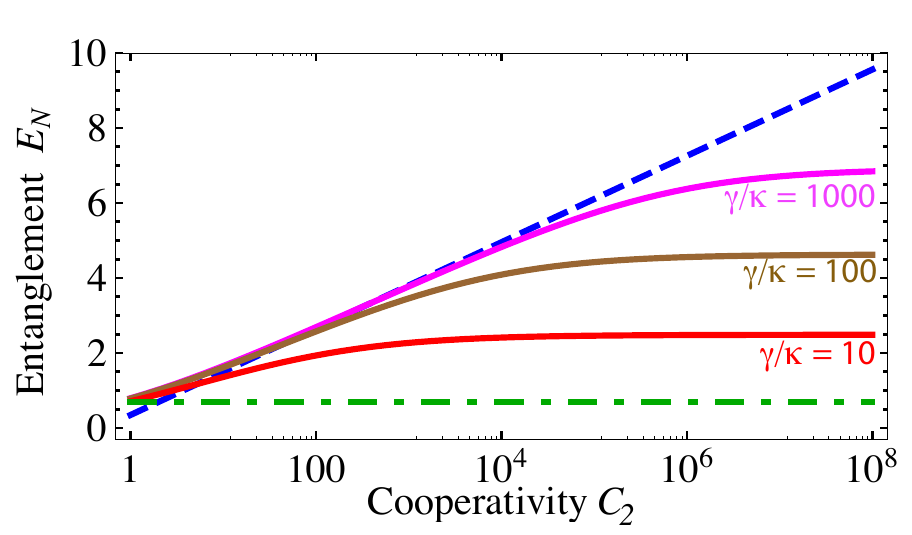}
\end{center}
\caption{Intra-cavity stationary entanglement (quantified by $E_N$) versus cooperativity $C_2$, where we use an optimized choice for $G_1$ (as given by Eq.~(\ref{eq:OptimalGTilde})), and have taken $\kappa_1 = \kappa_2$ and zero temperature.  The solid lines correspond to different choices of the damping ratio $\gamma / \kappa$ as indicated; increasing $\gamma / \kappa$ increases the amount that one can cool the delocalized $\bhA$ mode, and hence enhances entanglement.
For large $C_2$, these curves asymptote to the value in Eq.~(\ref{eq:OptimalEN1}).
The dashed blue line is the asymptotic expression of Eq.~(\ref{eq:OptimalEN2}).  The green dashed-dotted line indicates $E_N = \ln 2$ (the maximum $E_N$
possible with a two-mode squeezing interaction).}
\label{fig:Figure3}
\end{figure}


\textit{Maximizing entanglement--}  We now see that entanglement generation is more subtle than one might expect given the simple form of Eq.~(\ref{eq:Hint}). In particular, if $G_1$ is fixed, the amount of stationary entanglement is a {\it non-monotonic function} of the entangling-interaction strength $G_2$ 
(see Fig.~\ref{fig:Figure2}).
The dissipative entanglement mechanism discussed here directly explains this behaviour, as increasing $G_2$ has two opposing effects:  it not only
increases $r$ and the delocalization of the Bogoliubov modes (enhancing entanglement), but also increases the effective temperature of these modes.  
This latter effect is due both to an increase in the effective temperature of the cavity vacuum noise (c.f.~Eq.~(\ref{n00f})), and to a suppression of the cavity-cooling effect (as the effective coupling $\tilde{G}$ decreases with increasing $G_2$).      



The maximum entanglement is achieved by carefully balancing the opposing tendencies described above; without this optimization, the entanglement will remain small.
For fixed couplings $G_1, G_2$, one can optimize the entanglement as a function of the mechanical damping $\gamma$.  The maximum occurs at a {\it non-zero} dissipation strength, which at zero mechanical temperature 
and in the good-cavity limit is simply given by  $\gamma = 2 \tilde{G}$.
This value simply minimizes the occupancy of $\bhA$, and corresponds to a simple impedance matching condition (i.e.~the rate with which the $\bhA$ mode and mechanics exchange energy matches the rate at which the mechanics and its bath exchange energy). 

More relevant to experiment is to consider $\kappa$ and $\gamma$ fixed, and optimize the entanglement over coupling strength.  
Focusing on the most interesting regime where the cooperativity $C_{2}\equiv G_{2}^{2}/(\Gammam \kappa ) \gg 1$, and considering the 
good-cavity limit and zero temperature (the mechanical resonator is also cooled to vacuum by the 3rd cavity mode), we find that for fixed $G_2$, the optimal $G_1$ is given by
\begin{equation}
	G_1^{\rm opt} \approx G_2 + \sqrt{\frac{ \kappa \gamma}{8} }  \left(1+ \frac{2 \kappa}{\gamma} \right), 
	 \;\;\;
	\textrm{i.e. }
	\frac{\tilde{G}^{\rm opt}}{\gamma} \approx \left( \frac{C_2 \kappa^2}{2 \gamma^2} \right)^{1/4}. 
	\label{eq:OptimalGTilde}
\end{equation}
Note that for large $C_2$, this optimal value can easily correspond to a strong interaction $\tilde{G} > \kappa, \gamma$.
Thus, in this optimal regime, the effects of our ``engineered reservoir" (cold mechanical resonator) on the target cavity modes cannot be described by a Markovian dissipator in a master equation; 
this is in stark contrast to standard dissipation-by-entanglement schemes.  

For the optimal value of $G_1$ above, the entanglement takes simple forms in two relevant limits.  
If we hold $C_2$ fixed while taking the limit $\gamma / \kappa \rightarrow \infty$, we have (at zero temperature):
\begin{equation}
		E_{N}^{\mathrm{opt}}	\approx 	\frac{1}{2} \ln \left[   2C_{2}  \right]
		= 2 r - 2 \ln 2
	\label{eq:OptimalEN2}
\end{equation}
For large $C_2$, the entanglement is almost that of a two-mode squeezed vacuum (i.e.~$E_N = 2r$).  Alternatively, we could
hold the ratio $\gamma / \kappa$ fixed and let
$C_2 \rightarrow \infty$, we have (at zero temperature):
\begin{equation}
		E_{N}^{\mathrm{opt}}	\sim
		\ln \left(2 + \frac{\gamma}{\kappa}  + \mathcal{O} \left[ \frac{ (\gamma / \kappa)^2}{\sqrt{C_2}} \right]  \right)
	\label{eq:OptimalEN1}
\end{equation}
This is the strong-interaction limit, where the $\bhA$ mode hybridizes with the mechanical resonator.  The maximal cooling
of $\bhA$ is consequently set by the ratio $\gamma / \kappa$ (c.f.~Eq.~(\ref{eq:BetaACooling})).  The amount of entanglement
here increases monotonically from $\ln 2$ (the maximum possible with a coherent coupling) as this cooling factor is increased.

The behaviour of the stationary entanglement versus coupling strength is shown in Fig.~\ref{fig:Figure2}, where we have used parameters similar to those
achieved in recent state-of-the-art experiments on microwave-circuit optomechanical systems \cite{Teufel2011,Teufel2011b}.
We assume that a $\omegam = 10 \text{ MHz}$ mechanical resonator is first cavity cooled to near its ground state, with a final damping rate of
$\gamma = 0.8 \text{ MHz}$ (which is predominantly due to the cold optical damping used for the cooling).  By then optimally tuning the couplings to the target modes
$G_1, G_2$ to optimize the dissipative entanglement mechanism (while keeping them $\lesssim 2.2 \text{ MHz}$), one 
can obtain a relatively large $E_N \sim 2.1$.  This exceeds by an order-of-magnitude the intra-cavity entanglement obtained in previous studies of the same
system \cite{Vitali2011},  as well as the the maximum of $\ln 2$ possible with a coherent two-mode squeezing interaction.
If this entanglement was used for a teleportation experiment, the maximum possible fidelity would be $0.89$~\cite{Adesso2005,Vitali2008}; this reduces the error by a factor of three compared to 
what would be possible with $E_N = \ln 2$.  Fig.~\ref{fig:Figure2} also shows that large values of $E_N$ are possible
even when $\bar{n}_{\rm th, M} \neq 0$ and $\kappa_1 \neq \kappa_2$.

In Fig.~\ref{fig:Figure3} we show how the stationary entanglement grows to dramatically large values with $C_2$ for an optimized choice of $G_1$.
While the parameters needed for such $E_N$ may be out of reach in current-generation optomechanics experiments, they may
be more feasible by implementing a superconducting circuit realization of our scheme \cite{Devoret2010b,Deppe2012}. 

\textit{Conclusions--}  We have presented a general method for the dissipative generation of entanglement in a three-mode optomechanical system.  
The entanglement generated here could be verified by measuring the covariance matrix of the two target cavity modes using homodyne techniques (see e.g.~\cite{Vitali2007c}).
Alternatively, one could directly use the cavity output spectra at resonance to measure the occupancy of the $\bhA$ mode; verifying that it violates the Duan inequality
of Eq.~(\ref{eq:Duan}) would also confirm the generation of entanglement (see \cite{EPAPS2013a}).

We thank S. Chesi and L. Tian for useful conversations.  This work was supported by the DARPA ORCHID program under a grant from the AFOSR.  





\newpage

\begin{widetext}

\section{Supplemental information}

\subsection{System stability condition}

The linearized Heisenberg-Langevin equations for our three mode system are
given in Eqs.~(6) of the main text. If we drop the noise terms, these
equations take the form $\vec{v} = \mathbf{M} \cdot \vec{v}$ where $\vec{v}
= (\hat{b}, \hat{d}_1, \hat{d}_2^\dag)$ and $\mathbf{M}$ is a $3 \times 3$
matrix. For the system to be stable, we require that the eigenvalues of $%
\mathbf{M}$ all have a negative real part. This requirement leads to the
well-known Routh-Hurwitz stability conditions. In the simple case $\kappa_1
= \kappa_2$, the stability conditions reduce to the following necessary and
sufficient condition: %
%
\begin{equation}
\tilde{G}^{2}>-\frac{\kappa \gamma }{4}
\end{equation}%
Thus, if $\kappa_1 = \kappa_2$ and $G_2 \leq G_1$, the system is always
stable (regardless of the magnitude of the cavity and mechanical damping
rates).

In contrast, if $\kappa_1 \neq \kappa_2$, then the stability conditions are
modified such that the system can be unstable at $\tilde{G}=0$. While the
general form of the conditions is somewhat unwieldy, in the interesting
large cooperativity limit (i.e.~ $C_i \equiv G_i^2 / (\kappa_i \gamma) \gg 0$%
), they reduce to: 
\begin{equation}
\tilde{G}^{2} > \bar{C} \gamma \max \left[ \kappa_1 - \kappa_2, \frac{%
\kappa_2^2 - \kappa_1^2}{2\gamma+\kappa_1 + \kappa_2} \right]
\end{equation}
with $\bar{C} \equiv (G_1^2 + G_2^2) / (\gamma (\kappa_1 + \kappa_2))$. Note
that in the $\gamma \gg \kappa_1, \kappa_2$ limit of interest, this
condition is very sensitive to the sign of $\kappa_1 - \kappa_2$. On a
heuristic level, damping asymmetry leads to negative damping terms in the
equations of motion for the modes $\bhA, \bhB$. For $\kappa_1 < \kappa_2$
(and $\tilde{G} > 0$), it is $\bhA$ which experiences negative damping; this
is overwhelmed by the positive cold damping $\Gamma_{\mathrm{opt}}$ provided
by the interaction with the mechanics (c.f.~Eq.~(8) in the main text), and
hence there is no possibility of instability. In contrast, if $\kappa_1 >
\kappa_2$, it is the dark mode $\bhB$ which experiences negative damping; as
there is nothing to offset this, the system can become unstable.

\subsection{Definition of the logarithmic negativity}

In this paper, unless specified, we use the \textit{logarithmic negativity} $%
E_{N}$ to quantify the degree of entanglement; this quantity is a rigorous
entanglement monotone, and is zero for separable states. For two-mode
Gaussian states of sort realized by the two cavity modes $\hat{d}_{1},\hat{d}%
_{2}$ in our system, it can be calculated using the expression \cite%
{Vidal2002} 
\begin{equation}
E_{N}=\max [0,-\ln 2\eta ^{-}]
\end{equation}%
with%
\begin{equation}
\eta ^{-}=\frac{1}{\sqrt{2}}\sqrt{\Sigma -\sqrt{\Sigma ^{2}-4\det \mathbf{V}}%
}
\end{equation}%
and%
\begin{equation}
\Sigma =\det \mathbf{B}+\det \mathbf{B}^{\prime }-2\det \mathbf{C}.
\end{equation}%
Here $\mathbf{V}$ is the $4\times 4$ covariance matrix of the two modes of
interest, defined via $V_{jj^{\prime }}=\frac{1}{2}\left\langle \Delta \hat{%
\xi}_{j}\Delta \hat{\xi}_{j^{\prime }}+\Delta \hat{\xi}_{j^{\prime }}\Delta 
\hat{\xi}_{j}\right\rangle $, with $\Delta \hat{\xi}_{j}=\hat{\xi}%
_{j}-\langle \hat{\xi}_{j}\rangle $ and $\hat{\vec{\xi}}=\left\{ \hat{x}_{1},%
\hat{p}_{1},\hat{x}_{2},\hat{p}_{2}\right\} $. Here $\hat{x}_{i}=\left( \hat{%
d}_{i}+\hat{d}_{i}^{\dag }\right) /\sqrt{2}$ and $\hat{p}_{i}=-i\left( \hat{d%
}_{i}-\hat{d}_{i}^{\dag }\right) /\sqrt{2}$ and $\left[ \hat{d}_{i},\hat{d}%
_{j}^{\dag }\right] =\delta _{ij}$. The matrix $\mathbf{B}$, $\mathbf{B}%
^{\prime }$ and $\mathbf{C}$ are $2\times 2$ matrices related to the
covariance matrix $\mathbf{V}$ as%
\begin{equation}
\mathbf{V=}\left( 
\begin{array}{cc}
\mathbf{B} & \mathbf{C} \\ 
\mathbf{C}^{T} & \mathbf{B}^{\prime }%
\end{array}%
\right) .
\end{equation}

The evaluation of entanglement needs the full information of the covariance
matrix. In our main text, the relevant system consists of two cavity modes $%
\hat{d}_{1}$ and $\hat{d}_{2}$, or equivalently $\hat{\beta}_{A}$ and $\hat{%
\beta}_{B}$. Thus we also the correlation of $\hat{\beta}_{A}$ and $\hat{%
\beta}_{B}$, beside their occupancy (Eqs.~(7) and (8) in the main text). The
only non-zero correlator (again working in the good-cavity limit $\omega _{%
\mathrm{M}}\gg \kappa _{1},\kappa _{2}$ where the counter-rotating terms $%
\hat{H}_{CR}$ have negligible effect) is 
\begin{equation}
\left\langle \hat{\beta}_{A}\hat{\beta}_{B}\right\rangle =\frac{\kappa
\left( \kappa +\gamma \right) }{\Gamma _{\mathrm{opt}}\gamma +2\kappa \left(
\kappa +\gamma \right) }(\bar{n}_{\mathrm{th},1}+\bar{n}_{\mathrm{th}%
,2}+1)\sinh 2r.
\end{equation}

\subsection{Entanglement of mechanical resonators mediated by a cavity mode}

While the main text focuses on a 3-mode optomechanical system having two
cavity modes coupled to a single mechanical mode, our
entanglement-by-dissipation scheme is very general, and can be applied in
principle to any set of three parametrically-coupled bosonic modes. In
particular, it could be realized in a 3-mode optomechanical system where two
mechanical modes are coupled to a single cavity mode.  We analyze this case below.
Note that after this work was submitted, a paper by Tan et al. appeared which also analyzes 
dissipative entanglement of two mechanical resonators coupled to a cavity \cite{Tan2013};  unlike our analysis before, 
this work does not explicitly consider the effect of non-RWA terms, terms we find to be especially
problematic.

The starting
Hamiltonian is now 
\begin{equation}
\hat{H}=\omega _{\mathrm{cav}}\hat{a}^{\dagger }\hat{a}+\sum_{i=1,2}\left(
\omega _{i}\hat{b}_{i}^{\dag }\hat{b}_{i}+g_{i}\left( \hat{b}_{i}^{\dag }+%
\hat{b}_{i}\right) \hat{a}^{\dagger }\hat{a}\right) +\hat{H}_{\mathrm{diss}}.
\end{equation}%
Here $\hat{b}_{i}$ is the annihilation operator of the $i$-th mechanical
resonator (frequency $\omega _{i}$, damping rate $\gamma _{i}$), $\hat{a}$
is the annihilation operator of the auxiliary cavity mode (frequency $\omega
_{\mathrm{cav}}$, damping rate $\kappa $), and $g_{i}$ are the
optomechanical coupling strengths. $\hat{H}_{\mathrm{diss}}$ describes the
dissipation of each mode, as well as the driving of the cavity mode.

To achieve an entangling interaction, we will drive the cavity mode at two
frequencies $\omega _{d1}=\omega _{\mathrm{cav}}-\omega _{1}$ and $\omega
_{d2}=\omega _{\mathrm{cav}}+\omega _{2}$ (i.e.~the red(blue) sideband
associated with mechanical resonator $1$ ($2$)). We can then write the
cavity annihilation operator as $\hat{a}=\bar{a}(t)+\hat{d}e^{-i\omega _{%
\mathrm{cav}}t}$, where the classical cavity amplitude $\bar{a}(t)=\left( 
\bar{a}_{1}e^{i\omega _{1}t}+\bar{a}_{2}e^{-i\omega _{2}t}\right)
e^{-i\omega _{\mathrm{cav}}t}$ is obtained from the classical equations of
motion, with the sideband amplitudes $\bar{a}_{1,2}$ determined
independently by the two cavity drives. We take $|\bar{a}_{1,2}|\gg 1$ which
allows us to linearize the optomechanical interaction in the usual way
(i.e.~we drop interaction terms not enhanced by the classical cavity
amplitude). Working in an interaction picture with respect to $\hat{H}%
_{0}=\omega _{1}\hat{b}_{1}^{\dag }\hat{b}_{1}+\omega _{2}\hat{b}_{2}^{\dag }%
\hat{b}_{2}+\omega _{\mathrm{cav}}\hat{d}^{\dag }\hat{d}$, the linearized
Hamiltonian in the displaced frame can be written as $\hat{H}=\hat{H}_{%
\mathrm{int}}+\hat{H}_{\mathrm{CR}}+\hat{H}_{\mathrm{diss}}$ with 
\begin{equation}
\hat{H}_{\mathrm{int}}=G_{1}\left( \hat{d}^{\dag }\hat{b}_{1}+\hat{d}\hat{b}%
_{1}^{\dag }\right) +G_{2}\left( \hat{d}\hat{b}_{2}+\hat{d}^{\dag }\hat{b}%
_{2}^{\dag }\right) .  \label{eq:Hint}
\end{equation}%
Here $G_{i}=g_{i}\bar{a}_{i}$ (we take $g_{i}$, $\bar{a}_{i}$ to be positive
without loss of generality) and $\hat{H}_{\mathrm{CR}}$ describes
counter-rotating terms with an explicit time-dependence. Eq.~(\ref{eq:Hint})
takes the same form as Eq. (2) in the main text and can be used for
entangling $\hat{b}_{1}$ and $\hat{b}_{2}$ by reservoir engineering.

As discussed in the main text, the dissipative entanglement mechanism is
optimal when the mode acting as the engineered reservoir is both cold and
has a large damping rate compared to the target modes, as this allows
optimal cooling of the Bogoliubov mode $\bhA$. This occurs naturally here,
as the reservoir mode is a cavity mode, which given its higher frequency
(compared to mechanical modes) will be naturally closer to the ground state,
and which naturally has a large damping rate (i.e.~its damping $\kappa$ will
be much larger than the damping rates $\gamma$ of the target mechanical
modes, by a factor $\sim 10^{4}$ to $10^{5}$ in typical optomechanical
experiments). %
An additional benefit here is that since the mechanical damping rate $\gamma$
is small, the cooperativity of the system can reach a much larger value (in
the strong coupling limit, $C$ can reach $10^{6}$). Thus the entanglement
can easily reach $E_{N}\sim 7$ (as shown in Fig. 3), corresponding to
teleportation fidelity of $99.9\%$. In contrast, the
two-cavities-plus-mechanics scheme in the main text requires one to
optically damp the mechanical damping $\gamma$ to a large value.

While the above features are extremely attractive, the downside of this
scheme comes when one considers the non-resonant, counter-rotating terms.
For the setup discussed in the main text (two cavities, one mechanical
resonator), the counter-rotating terms only lead to a very small heating of
the coupled $\bhA$ mode, much like standard optomechanical cavity cooling
(c.f.~Eq.~(5) in the main text). In contrast, the situation here (two
mechanical resonators, one cavity) is more complex. In general there are two
sets of non-resonant, counter-rotating terms. The first are
\textquotedblleft diagonal terms", where, e.g., the cavity drive at $\omega
_{\mathrm{cav}}-\omega _{1}$ induces Stokes scattering with mechanical
resonator 1. In our rotating frame, these take the form: %
\begin{equation}
H_{\mathrm{CR}}^{A}=\left( \bar{a}_{1}^{\ast }g_{1}\hat{d}\hat{b}%
_{1}e^{-2i\omega _{1}t}+h.c.\right) +\left( \bar{a}_{2}^{\ast }g_{2}\hat{d}%
\hat{b}_{2}^{\dag }e^{2i\omega _{2}t}+h.c.\right) .
\end{equation}%
We also have non-resonant \textquotedblleft off-diagonal" terms where now,
e.g., the cavity drive at $\omega _{\mathrm{cav}}-\omega _{1}$ induces both
Stokes and anti-Stokes scattering involving mechanical resonator $2$. These
take the form: 
\begin{eqnarray}
H_{\mathrm{CR}}^{B} &=&\bar{a}_{1}^{\ast }g_{2}\left( e^{i\left( \omega
_{2}-\omega _{1}\right) t}\hat{d}\hat{b}_{2}^{\dag }+e^{-i\left( \omega
_{1}+\omega _{2}\right) t}\hat{d}\hat{b}_{2}\right)   \notag \\
&&+\bar{a}_{2}^{\ast }g_{1}\left( e^{i\left( \omega _{2}-\omega _{1}\right)
t}\hat{d}\hat{b}_{1}+e^{i\left( \omega _{1}+\omega _{2}\right) t}\hat{d}\hat{%
b}_{1}^{\dag }\right) +h.c.
\end{eqnarray}

To understand the effects of these terms, we re-express them in terms of the
Bogoluibov modes $\bhA, \bhB$ using Eq. (4) in the main text (with the
proviso that in these equations, $\hat{d}_{i}$ should be replaced with $\hat{%
b}_{i}$). We thus have: 
\begin{eqnarray}
H_{\mathrm{CR}}^{A} &=&\tilde{G}\hat{d}\hat{\beta}_{A}\left( \cosh
^{2}re^{-2i\omega _{1}t}-\sinh ^{2}re^{2i\omega _{2}t}\right)  \notag \\
&&+\tilde{G}\hat{d}\hat{\beta}_{B}^{\dag }\sinh r\cosh r\left( e^{2i\omega
_{2}t}-e^{-2i\omega _{1}t}\right) +h.c.
\end{eqnarray}%
and%
\begin{eqnarray}
H_{\mathrm{CR}}^{B} &=&\bar{a}_{1}^{\ast }g_{2}\left( e^{i\left( \omega
_{2}-\omega _{1}\right) t}\hat{d}\left( -\hat{\beta}_{A}\sinh r+\hat{\beta}%
_{B}^{\dag }\cosh r\right) +e^{-i\left( \omega _{1}+\omega _{2}\right) t}%
\hat{d}\left( -\hat{\beta}_{A}^{\dag }\sinh r+\hat{\beta}_{B}\cosh r\right)
\right)  \notag \\
&&+\bar{a}_{2}^{\ast }g_{1}\left( e^{i\left( \omega _{2}-\omega _{1}\right)
t}d\left( \hat{\beta}_{A}\cosh r-\hat{\beta}_{B}^{\dag }\sinh r\right)
+e^{i\left( \omega _{1}+\omega _{2}\right) t}\hat{d}\left( \hat{\beta}%
_{A}^{\dag }\cosh r-\hat{\beta}_{B}\sinh r\right) \right)  \notag \\
&&+h.c.
\end{eqnarray}

The crucial difference here is that unlike the the case discussed in the
main text (2 cavities, 1 mechanical resonator), the non-resonant terms when
written in the $\bhA,\bhB$ basis enter with factors of $G_{i}\cosh r$, $%
G_{i}\sinh r$ ($i=1,2$) (recall that $G_1 = \tilde{G} \cosh r $, $G_2 = \tilde{G}\sinh r $).
Their effects will thus be exponentially larger
than in the case considered in the main text, as in that case, the
counter-rotating terms only enter with a small prefactor $\propto \tilde{G}$
(compare against Eq.~(5) in the main text). As a result, if one wants a large squeeze
parameter $r$, one will need to be
extremely far into the good-cavity limit to suppress the deleterious effects
of the non-resonant terms.

To estimate how far one needs to be in the good cavity limit in this
2-mechanical resonator setup, note that the terms in $\hat{H}_{\mathrm{CR}%
}^{A},\hat{H}_{\mathrm{CR}}^{B}$ that contain $\hat{d}\hat{\beta}_{A}$, $%
\hat{d}\hat{\beta}_{B}$ (and their Hermitian conjugates) will cause negative
damping of the $\bhA,\bhB$ modes. If this negative damping becomes too large
(i.e.~larger than the intrinsic damping $\gamma $ of these modes), this can
cause instability. %
%
%
%
Insisting that the induced negative damping is less than $\gamma $ in the
large cooperativity, large-$r$ limit of interest (i.e.~where the
entanglement can be large) leads to the requirement: 
\begin{equation}
\frac{G_1^{2}e^{2r}}{\kappa }\left( \frac{\kappa }{\tilde{\omega}}%
\right) ^{2}\lesssim \gamma 
\end{equation}%
%
%
where $\tilde{\omega}=\min \left[ 2\omega _{1},2\omega _{2},|\omega
_{1}-\omega _{2}|\right] $. Assuming the optomechanical couplings $G_{1}$
has been chosen to optimize $E_{N}$ in the absence of counter-rotating terms
and with $G_{2}$ fixed (as per Eq.~(10) in the main text), we can re-write
this condition in terms of the cooperativity $C_{2}=G_{2}^{2}/(\kappa \gamma
)$ 
\begin{equation*}
\frac{\kappa }{\tilde{\omega}}\lesssim \left( \frac{1}{C_{2}}\right) ^{3/4}
\end{equation*}%
Thus, obtaining significant entanglement (which necessarily requires a large 
$C_{2}$) also requires one to be extremely deep into the good-cavity limit,
i.e.~$\omega _{\mathrm{M}}\gg \kappa $. For example, in order to achieve $%
E_{N}\sim 5$, the ratio $\kappa /\tilde{\omega}$ has to be smaller than $%
10^{-2}$, a value that would be extremely challenging for current
optomechanics experiments. Microwave cavity experiments have achieved a
ratio $\omega _{\mathrm{M}}/\kappa \sim 50$ \cite{Teufel2011b}; with such
numbers, an $E_{N}\gtrsim 1.5$ could be possible. This is already more than
a factor of two greater than the maximal entanglement $E_{N}=\ln 2$ possible
with a direct two-mode squeezing interaction.

%
%
%
%
%
%


\subsection{The bound of the Bogoliubov mode occupancy by the generalized Duan
inequality}

Using the definition of Bogoliubov mode Eq. (4) in the main text, one gets%
\begin{eqnarray}
\left\langle \left\{ \hat{\beta}_{A}^{\dag },\hat{\beta}_{A}\right\}
\right\rangle &\equiv &1+2\tilde{n}_{A}  \notag \\
&=&V_{11}\cosh ^{2}r+V_{22}\sinh ^{2}r+\left( C_{12}+C_{12}^{\ast }\right)
\sinh r\cosh r,
\end{eqnarray}%
with $V_{11}\equiv \left\langle \left\{ \hat{d}_{1}^{\dag },\hat{d}%
_{1}\right\} \right\rangle $, $V_{22}\equiv \left\langle \left\{ \hat{d}%
_{2}^{\dag },\hat{d}_{2}\right\} \right\rangle $ and $C_{12}\equiv
\left\langle \left\{ \hat{d}_{1},\hat{d}_{2}\right\} \right\rangle $ where $%
\left\{ \cdots \right\} $ denotes anti-commutator.

According to generalized Duan inequality \cite{Duan2000}, any separable
(non-entangled state) will necessarily satisfy 
\begin{equation}
D\equiv \left\langle \left( a\hat{x}_{1}+\frac{1}{a}\hat{x}_{2}\right)
^{2}\right\rangle +\left\langle \left( a\hat{p}_{1}-\frac{1}{a}\hat{p}%
_{2}\right) ^{2}\right\rangle \geq \frac{1}{a^{2}}+a^{2},
\end{equation}%
where $\hat{x}_{i}=\left( \hat{d}_{i}+\hat{d}_{i}^{\dag }\right) /\sqrt{2}$
and $\hat{p}_{i}=-i\left( \hat{d}_{i}-\hat{d}_{i}^{\dag }\right) /\sqrt{2}$
and $a$ is an arbitrary nonzero real number.

If we set $a^{2}=\coth r$, we can easily verify that 
\begin{equation}
1+2\tilde{n}_{A}=\frac{D}{2}\sinh 2r.
\end{equation}%
For this choice of $a$, using the Duan bound on $D$ thus yields that for any
separable state, the occupancy of the $\bhA$ mode must satisfy: 
\begin{equation}
\tilde{n}_{A}\geq \sinh ^{2}r.
\end{equation}
Note from Eq.~(7) in the main text that at zero temperature at zero
optomechanical coupling, $\tilde{n}_A$ is exactly $\sinh^2 r$, indicating no
violation of the Duan bound (consistent with no entanglement in this case).
However, any amount of cooling (as per Eq.~(8)) in the main text will then
lower the value of $\tilde{n}_A$, leading to a violation of the Duan bound
and hence entanglement of the two target cavity modes.


\subsection{Cooling one Bogoliubov mode vs. cooling two Bogoliubov modes}

As explained after Eq.~(4) of the main text, if both Bogoliubov modes ($\bhA$
and $\bhB$) are cooled to vacuum, a two-mode squeezed vacuum of $\hat{d}_{1}$
and $\hat{d}_{2}$ can be achieved. Two-mode squeezed vacuum is a
highly-entangled state with entanglement (evaluated by log negativity) $%
E_{N}=2r$. This is the central idea lying in previous works of dissipation
generated entanglement (see. e.g. Ref [21-22] in the main text). However, in
our setup, one of the Bogoliubov modes $\bhB$ is decoupled from the
intermediate mode (cf. Eq. (5)) which makes it impossible to be cooled by
laser-cooling mechanism. Thus the ideal steady-state to be achieved with our
protocol, $\hat{\rho}_{0}$, is characterized by the following correlations%
\begin{equation}
\left\langle \bhA^{\dag }\bhA\right\rangle =0,\left\langle \bhB^{\dag }\bhB%
\right\rangle =\sinh ^{2}r,\left\langle \bhB\bhA\right\rangle =0
\end{equation}%
where the occupany of the $\bhB$ mode increases exponentially with the
squeezing parameter $r$. The corresponding covariance matrix of the state $%
\hat{\rho}_{0}$ in the basis of $\left\{ \hat{x}_{1},\hat{p}_{1,}\hat{x}_{2},%
\hat{p}_{2}\right\} $ (where $\hat{x}_{i}=(\hat{d}_{i}+\hat{d}_{i}^{\dag })/%
\sqrt{2}$ and $\hat{p}_{i}=-i(\hat{d}_{i}-\hat{d}_{i}^{\dag })/\sqrt{2}$)\
is 
\begin{equation}
\mathbf{V}(\hat{\rho}_{0})=\frac{1}{2}\left( 
\begin{array}{cccc}
3+\cosh 4r & 0 & -8\cosh ^{3}r\sinh r & 0 \\ 
0 & 3+\cosh 4r & 0 & 8\cosh ^{3}r\sinh r \\ 
-8\cosh ^{3}r\sinh r & 0 & 4\cosh 2r+\cosh 4r-1 & 0 \\ 
0 & 8\cosh ^{3}r\sinh r & 0 & 4\cosh 2r+\cosh 4r-1%
\end{array}%
\right) .  \label{v2msts}
\end{equation}%
The overlap between this state and an arbitrary two mode squeezed vacuum $|%
\tilde{r},\theta \rangle =\exp \left[ \tilde{r}e^{i\theta }\hat{d}_{1}\hat{d}%
_{2}-h.c.\right] |0,0\rangle $ (where $|0,0\rangle $ is the vacuum state of
the modes $\hat{d}_{1}$ and $\hat{d}_{2}$) can be evaluated as 
\begin{equation}
F\equiv \mathrm{Tr}\left( \hat{\rho}_{0}\hat{\rho}_{\mathrm{TMSV}}\right) =%
\frac{4}{\cosh ^{2}r}\frac{1}{2+\cosh \left( r-\tilde{r}\right) +\cosh
\left( r+\tilde{r}\right) -2\cos \theta \sinh 2r\sinh 2\tilde{r}},
\end{equation}%
where $\hat{\rho}_{\mathrm{TMSV}}=|\tilde{r},\theta \rangle \langle \tilde{r}%
,\theta |$ is the density matrix of the 2-mode squeezed vacuum. For large
squeezing $r\gg 1$, the overlap between the two states is negligible $F\sim
\exp \left( -6r\right) \rightarrow 0$, as mentioned after Eq.~(9) in the
main text.

Of course the cooling of $\bhB$ mode can also be achieved by introducing a
second intermediate mode and two-tone drives on the cavity modes. However,
we find this extra complication is not necessary for our purpose. In fact,
although our scheme is insufficient to generate a 2-mode squeezed vacuum,
surprisingly it is sufficient to generate a steady state with significant
entanglement $E_{N}=2r-\ln 2$, which is almost as good as a 2-mode squeezed
vacuum in the large $r$ limit. To understand the physical meaning of the
state $\hat{\rho}_0$, we can compare Eq.~(\ref{v2msts}) with the covariance
matrix of a 2-mode squeezed thermal state~\cite{Marian2003} 
\begin{equation}
\rho _{\mathrm{TMST}}=\tilde{S}\left( r\right) \rho _{1,\mathrm{th}}\otimes
\rho _{2,\mathrm{th}}\tilde{S}^\dag\left( r\right)
\end{equation}%
with 
\begin{equation}
\rho _{i,\mathrm{th}}=\sum_{n_{i}=1}^{\infty}\frac{(\bar{n}^\mathrm{e}%
_{i})^{n_{i}}}{\left( 1+\bar{n}^\mathrm{e}_{i}\right) ^{n_{i}+1}}\left\vert
n_{i}\right\rangle \left\langle n_{i}\right\vert
\end{equation}%
the thermal state of mode $i$ ($i=1,2$) with average thermal occupancy $\bar{%
n}^\mathrm{e}_{i}$. Interestingly, we find that, Eq.~(\ref{v2msts}) can be
reproduced by setting $\bar{n}^\mathrm{e}_{1}=0$ and $\bar{n}^\mathrm{e}%
_{2}=\sinh ^{2}r$. This shows the state obtained in our work is a two-mode
squeezed thermal state with the two modes at different effective
temperatures.


\subsection{Measuring entanglement via the $\hat{\protect\beta}_A$ occupancy}

As shown in the previous section, if one can verify that the occupancy of
the $\bhA$ mode has been reduced below the value $\sinh ^{2}r$ (where $r=%
\mathrm{arctanh}\,G_{2}/G_{1}$), then one has violated the Duan inequality,
indicating that cavities $1$ and $2$ are necessarily entangled. We now show
that this mode occupancy can be directly obtained from the output spectra of
the two cavities. Consider the simplest case where all dissipative baths are
at zero temperature, and where we are deep in the good-cavity limit; for
concreteness, also focus on the output spectrum of cavity $1$. 
The solution to the Heisenberg-Langevin equation of motion for $\hat{d}_{1}$
may be written: 
\begin{equation*}
\hat{d}_{1}[\omega ]=\chi _{1}[\omega ]\left( -iG_{1}\chi _{M}[\omega
]\left( -i\tilde{G}\bhA[\omega]-\sqrt{\gamma }\hat{b}_{\mathrm{in}}[\omega
]\right) -\sqrt{\kappa }\hat{d}_{1,\mathrm{in}}[\omega ]\right) 
\end{equation*}%
where the susceptibilities are: 
\begin{eqnarray}
\chi _{1}[\omega ] &=&\frac{1}{-i(\omega -\omega _{\mathrm{M}})+\kappa /2} \\
\chi _{M}[\omega ] &=&\frac{1}{-i(\omega -\omega _{\mathrm{M}})+\gamma /2}
\end{eqnarray}%
Further, the output field from cavity $1$ $\hat{d}_{1,\mathrm{out}}$ is
given by the standard input-output relation 
\begin{equation*}
\hat{d}_{1,\mathrm{out}}=\hat{d}_{1,\mathrm{in}}+\sqrt{\kappa }\hat{d}_{1},
\end{equation*}%
and its power spectral density defined as 
\begin{equation*}
S_{1}[\omega ]=\int dte^{i\omega t}\langle \hat{d}_{1,\mathrm{out}}^{\dag
}(t)\hat{d}_{1,\mathrm{out}}(0)\rangle .
\end{equation*}%
Note that the above equations (like the equations in the main text) are in
the rotating frame set by the cavity drive frequencies; hence, $\omega
=\omega _{\mathrm{M}}$ corresponds to the cavity resonance.

We see immediately that the only contribution to $S_1[\omega]$ will be from
the non-zero occupancy of $\bhA$. We focus for simplicity on the regime
where the effective coupling $\tilde{G}$ is less than $\gamma$, a regime
where large entanglement is possible but the $\bhA$ mode does not hybridize
with the mechanical resonator; the entanglement in this regime is described
by Eq.~(11) of the main text. In this regime, the total damping rate of the $%
\bhA$ mode will be $\kappa + \Gamma_{\mathrm{opt}}$, where the effective
cold damping rate due to the mechanics is $\Gamma_{\mathrm{opt}} = 4 \tilde{G%
}^2 / \gamma$ (c.f. Eq.~(8) in the main text). A straightforward calculation
then shows that 
\begin{eqnarray}
S_1[\omega_{\mathrm{M}}] = 16 C_1 \frac{4 \tilde{C} }{1 + 4 \tilde{C}}
\langle \bhA^\dagger \bhA \rangle
\end{eqnarray}
where the co-operativities $C_1$ and $\tilde{C}$ are defined as $C_1 = G_1^2
/ (\kappa \gamma)$, $\tilde{C} = \tilde{G}^2 / ( \kappa \gamma)$. We thus
see that the output spectrum of cavity $1$ at resonance (i.e.~the number of
photons leaving the cavity at its resonance frequency) gives a direct
measure of the occupancy of the coupled Bogoluibov mode $\bhA$.

\end{widetext}

\end{document}